\documentclass[12pt]{article}
\usepackage{amssymb}
\usepackage{a4wide}
\usepackage{slashed}
\begin{document}
\pagestyle{empty}
\begin{flushright}
\small
SISSA 84/01/EP\\
{\bf hep-th/0111031}\\
%June  $16$th, $1998$
\normalsize
\end{flushright}
\begin{center}
{\large\bf A Small Note on PP-Wave Vacua in 6 and 5 Dimensions}\\[.5cm]
{\bf P.~Meessen}\\[.2cm]%\footnote{E-mail: {\tt meessen@sissa.it}}\\[.2cm]
{\em International School for Advanced Studies (SISSA/ISAS)\\
Via Beirut 2-4, 34014 Trieste, Italy}\\[.5cm]
{\bf abstract}\\
\begin{quote}
{\small 
We discuss Kowalski-Glikman type pp-wave solutions with unbroken supersymmetry in
6 and 5 dimensional supergravity theories.
}
\end{quote}
\end{center}
\vspace{.5cm}
%%%%%%%%%%%%%%%%%%%%%%%%%%%%%%%%%%%%%%%%%%%%%%%%%%%%%%%%%%%%%%%%%%%%%%%%%%%
\pagestyle{plain}
%%%%%%%%%%%%%%%%%%%%%%%%%%%%%%%%%%%%%%%%%%%%%%%%%%%%%%%%%%%%%%%%%%%%%%%%%%
In this small note we want to discuss pp-wave solutions
with unbroken supersymmetry, the so-called Kowalski-Glikman solutions,
in lower dimensional supergravity theories. 
The known KG solutions \cite{art:K_1,art:blau,art:K_3} consist of some covariantly constant
fieldstrength and a metric which has the form of a pp-wave, {\em i.e.}
\begin{equation}
\label{eq:wave_metric}
ds_{d}^{2} \;=\; 2du\left( dv\,+\, Adu\right)\;+\; dx^{i}dx_{i} \; ,
\end{equation}
where $A= x^{i}A_{ij}x^{j}$ ($i,j=1\ldots d-2$). For this metric, the only non-vanishing component
of the Ricci curvature is $R_{uu}= 2\eta^{ij}A_{ij}$ and by introducing light-cone coordinates
in the tangent-space we find that the only non-vanishing component of the spin-connection is
$\slashed{\omega}_{\mu}= 2\delta_{u\mu}\partial_{i}A\gamma^{i+}$.\footnote{We use the mostly minus
signature for the metric, the $\gamma$ matrices satisfy $\{ \gamma^{\mu},\gamma^{\nu}\} =2g^{\mu\nu}$
and the covariant derivative on spinors is taken to be 
$\nabla\epsilon =d\epsilon -4^{-1}\slashed{\omega}\epsilon$. We also introduce the light-cone combinations
$\gamma^{+}=\gamma^{u}$ and $\gamma^{-}=\gamma^{v}+A\gamma^{u}$, which satisfy 
$\{ \gamma^{+},\gamma^{-}\} = 2$ and $\{\gamma^{\pm},\gamma^{\pm}\} = \{\gamma^{\pm},\gamma^{i}\} =0$.}
\par
The basic problem one is faced with when looking for non-trivial solutions of supergravity theories
which preserve all supersymmetry are the dilatino equations, since they are algebraic in
nature. Clearly, the easiest way to avoid such trouble is by not having dilatino equations
in the first place. Sometimes however, one can make use of special properties of the
dilatino equations in order to find non-trivial vacua. The first example is $d=10$, $N=1$
supergravity where one can use the chirality of the theory \cite{art:K_2} in order to
find a solution which preserves all supersymmetry. Another example is type IIB supergravity,
where one can find such such solutions, notably the $aDS_{5}\otimes S^{5}$ solution and the
Kowalski-Glikman type solution presented in \cite{art:blau}, with RR 5-form flux since the dilatino equation
does not contain a contribution of the 5-form fieldstrength \cite{art:S}. The fact that the type IIB dilatino
variation does not depend on the 5-form fieldstrength is however due to the fact that it is
self-dual and that the spinors are chiral.
One is therefore tempted to say that non-trivial solutions with unbroken supersymmetry exist
whenever there are no dilatinos or when the theory is chiral and it is these kind of theories
we are going to examine.
\par
There are not many supergravity theories that are chiral or have no dilatinos, so that the investigation
of the existence of KG solutions is rather limited. The highest dimensional possibilities have
already been presented in the literature, namely by Kowalski-Glikman in the case of M-theory \cite{art:K_1}
and by Blau {\em et. al.} for type IIB \cite{art:blau}. The next on the list is $N=1$ $d=10$ supergravity. 
Such an investigation was carried out by Kowalski-Glikman \cite{art:K_2} who showed that the solution
is not of the pp-wave type, but rather has geometry $aDS_{3}\otimes S^{3}\otimes\mathbb{E}_{4}$.
A similar analysis was performed on the $N=2$ $d=4$ supergravity \cite{art:K_3} showing that the 
only supersymmetric solutions are the Robinson-Bertotti and the KG solution.
This means that the only remaining candidates are $d=6$ $(2,0)$ or $(4,0)$ supergravity and $d=5$ $N=2$ supergravity. 
Although $N=1$ $d=4$ supergravity matches the profile, it can be discarded since
the integrability condition for the Killing spinor equation implies that the space must be 
Riemann flat, {\em i.e.} Minkowski.
%%%%%%%%%%%%%%%%%%%%%%%%%%%%%%%%%%%%%%%%%%%%%%%%%%%%%%%%%%%%%%%%%%%%%%%%%%%%%%%%%%%%%%%%%%%%%%%%%%%%
%
%  Application to d=6 N=2 suegra
%
%%%%%%%%%%%%%%%%%%%%%%%%%%%%%%%%%%%%%%%%%%%%%%%%%%%%%%%%%%%%%%%%%%%%%%%%%%%%%%%%%%%%%%%%%%%%%%%%%%%%
\par
The $d=6$ $(2,0)$ supergravity comprises of the graviton, $e_{\mu}^{a}$, two symplectic Majorana-Weyl Rarita-Schwinger
fields, combined into the $USp(2)$ vector $\Psi_{\mu}$, and a 2-form $B$ whose fieldstrength, $H=dB$, is self-dual. 
As such one is faced with the same problem as
in type IIB supergravity. A Lorentz invariant action can however be written down by introducing
a Lagrange multiplier field \cite{art:KM}, by writing a non-self-dual action as in \cite{art:BHO} or by 
adding an anti-symmetric tensor multiplet \cite{art:MS}. However, using the conventions of \cite{art:NS} 
the equation of motion for the metric reads
\begin{equation}
\label{eq:EOM_d=6_N=2}
  R_{\mu\nu} \;=\; \textstyle{\frac{1}{4}}\ H_{\mu\kappa\rho}H_{\nu}{}^{\kappa\rho} \; .
\end{equation}
Choosing the self-dual Ansatz $H= \lambda du\wedge\left(dx^{1}dx^{2}+dx^{3}dx^{4}\right)$ the above equation is 
solved by choosing $2\eta^{ij}A_{ij}=-\lambda^{2}$.
\par
The Killing spinor equations in this case read
\begin{equation}
\label{eq:d=6_susy_var}
 0\;=\; \delta\Psi_{\mu} \;=\; \nabla_{\mu}\epsilon \,-\, 
        \textstyle{\frac{1}{8\cdot 3!}}\slashed{H}\gamma_{\mu}\epsilon \; .
\end{equation}
By observing that due to the self-duality of $H$ we have $\slashed{H}\gamma_{\mu}\epsilon=
2\cdot 3!\lambda\gamma^{+}\gamma^{12}\gamma_{\mu}\epsilon$, we can see that the Eq. (\ref{eq:d=6_susy_var})
is automatically satisfied in the $v$ direction. The equations in the $i$ direction reads
\begin{equation}
 0\;=\; \partial_{i}\epsilon \;-\; \Omega_{i}\epsilon \hspace{.3cm}:\hspace{.3cm} 
        \Omega_{i} \;=\; \textstyle{\frac{1}{4}}\lambda\gamma^{+}\gamma^{12}\gamma_{i} \; .
\end{equation}
{}Following \cite{art:blau,art:FP} these equations, since $\Omega_{i}\Omega_{j}=0$, are solved by
$\epsilon \;=\; \left( 1+x^{i}\Omega_{i}\right)\xi (u)$.
In the $u$-direction Eq. (\ref{eq:d=6_susy_var}) reduces to
\begin{equation}
\partial_{u}\epsilon\,-\, x^{i}A_{ij}\gamma^{j}\gamma^{+}\epsilon \,-\, \Omega^{-}\epsilon \;=\; 0 \; , 
\end{equation}
where the combination $\Omega^{-}= \textstyle{\frac{1}{4}}\lambda\gamma^{+}\gamma^{12}\gamma^{-}$ was used.
By making the Ansatz $\xi = \exp\left( \Omega^{-}\ u\right)\epsilon_{0}$, with $\epsilon_{0}$ an unconstrained
constant symplectic-MW spinor, all $x$-independent terms are canceled leaving
\begin{equation}
  \label{eq:d=6_cond_A}
  x^{i}A_{ij}\gamma^{j}\gamma^{+}\epsilon_{0} \;=\; -\textstyle{\frac{1}{8}}\lambda^{2} x^{i}\gamma_{i}\ \gamma^{+}\epsilon_{0} \; .
\end{equation}
This equation is readily solved by $A_{ij}=-\textstyle{\frac{1}{8}}\lambda^{2}\eta_{ij}$, which is compatible
with the equations of motion.
%%%%%%%%%%%%%%%%%%%%%%%%%%%%%%%%%%%%%%%%%%%%%%%%%%%%%%%%%%%%%%%%%%%%%%%%%%%%%%%%%%%%%%%%%%%%%%%%%%%%%
%
%   Note on D=6 (4,0) theory
%
%%%%%%%%%%%%%%%%%%%%%%%%%%%%%%%%%%%%%%%%%%%%%%%%%%%%%%%%%%%%%%%%%%%%%%%%%%%%%%%%%%%%%%%%%%%%%%%%%%%%%
\par
The $d=6$ $(4,0)$ supergravity is invariant under global $USp(4)\sim SO(5)$ (See Ref. \cite{art:romans} and refs. therein) 
and its field content is a Sechsbein, $e_{\mu}{}^{a}$, 
4 symplectic MW spinor, which are combined into an $SO(5)$ vector $\Psi_{\mu}$, and 5 2-forms $B^{I}$ which 
have self-dual fieldstrengths and transform as a vector under $SO(5)$. The equations of motion and the 
supersymmetry transformation are
\begin{eqnarray}
0 &=& R_{\mu\nu} \;-\; \textstyle{\frac{1}{4}}\ H_{\mu\kappa\rho}^{I}H_{\nu}{}^{\kappa\rho\ I} \; , \nonumber \\
0 &=& \delta\Psi_{\mu} \;=\; \nabla_{\mu}\epsilon \;-\; 
        \textstyle{\frac{1}{8\cdot 3!}}\slashed{H}\gamma_{\mu}\Gamma^{I}\epsilon \; ,                      
\end{eqnarray} 
where the $\Gamma$'s belong to the 5-dimensional Euclidean Clifford algebra. The
$(2,0)$ solution can be embedded into the $(4,0)$ theory by taking only the $I=1$ component to be different
from zero. The calculations are just the same as in the $(2,0)$ case, the only difference being that 
every $\Omega$ has to be multiplied by $\Gamma^{1}$. The result is however the same: $(4,0)$ supergravity
admits a KG-type wave solution that breaks no supersymmetry whatsoever. 
%To put it differently: it works but I don't want to explain it again.
%%%%%%%%%%%%%%%%%%%%%%%%%%%%%%%%%%%%%%%%%%%%%%%%%%%%%%%%%%%%%%%%%%%%%%%%%%%%%%%%%%%%%%%%%%%%%%%%%%%%%
%
%    Application to N=2 D=5 supergravity
%
%%%%%%%%%%%%%%%%%%%%%%%%%%%%%%%%%%%%%%%%%%%%%%%%%%%%%%%%%%%%%%%%%%%%%%%%%%%%%%%%%%%%%%%%%%%%%%%%%%%%%
\par
The last on the short-list is $d=5$ $N=2$ supergravity \cite{art:cremmer}.
Its field content consists of a F\"unfbein, $e_{\mu}{}^{a}$,
two symplectic Majorana Rarita-Schwinger fields $\Psi_{\mu}$
and a vector $V_{\mu}$ whose fieldstrength will be taken to be
\begin{equation}
 F \;=\; du\wedge\ \lambda_{i}dx^{i} \; .
\end{equation}
Since the chosen form for the field strength is, given the metric (\ref{eq:wave_metric}),
covariantly constant and has an overall dependence on the differential $du$ the equation
of motion for the vector field is automatically satisfied. The equation of motion for the
metric reads
\begin{equation}
 R_{\mu\nu} \;=\; \textstyle{\frac{1}{2}}\left[
                     F_{\mu\kappa}F_{\nu}{}^{\kappa}\,-\,
                     \textstyle{\frac{1}{6}}g_{\mu\nu}F^{2}
                  \right] \; ,
\end{equation}
and leads to the condition $4\eta^{ij}A_{ij}=\lambda_{i}\lambda^{i}$. 
The supersymmetry variation of the gravitino reads \cite{art:cremmer}.
\begin{equation}
 0\;=\; \delta\Psi_{\mu}\;=\;
 \nabla_{\mu}\epsilon \,+\,
 \textstyle{\frac{1}{8\sqrt{3}}} \slashed{F}\gamma_{\mu}\epsilon \,-\,
 \textstyle{\frac{1}{4\sqrt{3}}} F_{\mu\nu}\gamma^{\nu}\epsilon \; .
\end{equation}
The analysis is completely analogous to the one for the $(2,0)$ theory,
but for the definitions
\begin{equation}
\Omega_{i} \;=\; -\textstyle{\frac{1}{4\sqrt{3}}}\gamma^{+}
                 \left[
                     \lambda_{j}\gamma^{j}\gamma_{i}+\lambda_{i}
                 \right] \; ,\;
\Omega^{-} \;=\; \textstyle{\frac{1}{4\sqrt{3}}}\lambda_{i}\gamma^{i}
               \left(
                  \gamma^{+}\gamma^{-}+1
               \right) \; .
\end{equation}
One finds that the analogue condition to Eq. (\ref{eq:d=6_cond_A}) reads
\begin{equation}
 x^{i}A_{ij}\gamma^{j}\gamma^{+}\epsilon_{0} \;=\; x^{i}\left[
                                \Omega_{i},\Omega^{-}
                                  \right] \epsilon_{0} \; ,
\end{equation}
where, as before, $\epsilon_{0}$ is an unconstrained symplectic Majorana spinor.
After some $\gamma$-manipulations, one finds the matrix $A$ to be
\begin{equation}
 A_{ij} \;=\; \textstyle{\frac{1}{24}}\left\{ \ 3\lambda_{i}\lambda_{j}\,+\,
                                                \lambda_{l}\lambda^{l}\eta_{ij}
                                      \ \right\} \; ,
\end{equation}
which is compatible with the equations of motion.
Of course we could make use of the $SO(3)$ invariance to put $\lambda_{2,3}=0$
and we find $A$ to be proportional to $diag(4,1,1)$.
\par
In this small note we have presented Kowalski-Glikman type solutions, solutions
that do not break any supersymmetry and that look like pp-waves, in 
in chiral 6-dimensional supergravities and in $N=2$ $d=5$ supergravity.
Since we used rather restrictive criteria, absence of dilatinos or
chirality, it would be nice to consider other theories and see whether
they admit to KG solutions. Work in this direction is in progress.
%%%%%%%%%%%%%%%%%%%%%%%%%%%%%%%%%%%%%%%%%%%%%%%%%%%%%%%%%%%%%%%%%%%%%%%%%%%%
%
%   de bedankjes...
%
%%%%%%%%%%%%%%%%%%%%%%%%%%%%%%%%%%%%%%%%%%%%%%%%%%%%%%%%%%%%%%%%%%%%%%%%%%%%
\section*{Acknowledgments}
The author would like to thank T. Ort\'{\i}n for encouragement.
This work was supported in part by the E.U. RTN program HPRN-CT-2000-00148.
%%%%%%%%%%%%%%%%%%%%%%%%%%%%%%%%%%%%%%%%%%%%%%%%%%%%%%%%%%%%%%%%%%%%%%%%%%%%
%
%    Referenties...
%
%%%%%%%%%%%%%%%%%%%%%%%%%%%%%%%%%%%%%%%%%%%%%%%%%%%%%%%%%%%%%%%%%%%%%%%%%%%

%%%%%%%%%%%%%%%%%%%%%%%%%%%%%%%%%%%%%%%%%%%%%%%%%%%%%%%%%%%%%%%%%%%%%%%%%%
%
%   Alea Iacta Est!
%
%%%%%%%%%%%%%%%%%%%%%%%%%%%%%%%%%%%%%%%%%%%%%%%%%%%%%%%%%%%%%%%%%%%%%%%%%%
\end{document}